# Quantum yield enhancement in BDMO-PPV

M. J. Deutsch and H. Park*
Department of Physics and Astronomy
Texas A&M University-Commerce, Commerce, Texas 75428, USA
*Corresponding author: heungman.park@tamuc.edu

**Abstract**
Poly-2,5-bis(3′,7′-dimethyloctyloxy)-1,4-phenylenevinylene (BDMO-PPV) is a photoluminescent semiconducting polymer related to others in the well-studied polyphenylene vinylene family such as poly-2-methoxy-5-(2-ethylhexyloxy)-1,4-phenylenevinylene (MEH-PPV). These materials are known for their instability, degradation, and low efficiency in device operations. We report increased internal and external quantum yield in BDMO-PPV during continuous illumination photodegradation through variations in absorbance, scattering, reflectance, and transmittance of the BDMO-PPV solution. We propose the creation of a highly emissive intermediate photoluminescent state responsible for these increased quantum yields.

## 1. Introduction

Semiconducting polymers have many unusual material properties as well as numerous uses in novel organic electronics. The conjugation of π-orbitals along the polymer backbone produces excellent optoelectronic properties useful for organic light-emitting diodes and solar cells. However, there are still some barriers to commercial usage, most notably the instability of organic semiconductors when exposed to oxygen and water in ambient conditions [1–14].

Poly[2,5-bis(3′,7′-dimethyloctyloxy)-1,4-phenylenevinylene] (BDMO-PPV) is a semiconducting polymer with chemical structure (Fig. 1) similar to other well-studied polymers such as poly[2-methoxy-5-(3',7'-dimethyloctyloxy)-1,4-phenylenevinylene] (MDMO-PPV), and poly[2-methoxy-5-(2-ethylhexyloxy)-1,4-phenylenevinylene] (MEH-PPV). Photoluminescence in these polymers is caused by exciting the electrons in the conjoined π-orbitals of the backbone and the side chains. The difference in side-chain structures between different poly-phenylenevinylene (PPV) derivatives can result in different optoelectronic material properties [15–19].

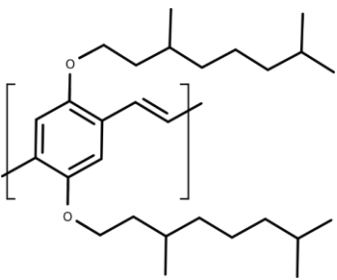

Figure 1 - BDMO-PPV chemical structure

The poly-phenylenevinylene family of semiconducting polymers has been shown to rapidly degrade in both solid thin films and solution suspensions. The most common of these processes is photobleaching or photo-oxidation due to the diffusion of oxygen into the polymer system [20–23]. The well-studied oxidation of the vinylene backbone results in chain scission and an overall reduction in conjugation length [22,24–29]. The degradation of thin films is more well-studied as thin films have applications in organic electronic device manufacturing. Studies of poly-phenylenevinylenes are comprised of either single-molecule spectroscopy or spectroscopy of an ensemble in solution [30,31].

As solution conformation can be preserved in films, understanding the solution dynamics of polymers is essential to understand thin film properties. Time-resolved fluorescence is usually used to study the short degradation of fluorescence emission, however, more research on continuous excitation of poly-phenylenevinylenes is desired. BDMO-PPV has been characterized in films and has had preliminary studies done on its photoluminescence spectra and its use in OLEDs [32–35]. Research examining the effects of long-term illumination and emission of BDMO-PPV solutions is lacking. Enhanced photoluminescence emission has been previously reported in films and solutions of MEH-PPV [36–39]. Previously, quantum yield enhancement was shown to occur in oxygen-poor environments either in a vacuum or in a nitrogen atmosphere [40,41]. Some quantum yield manipulation has been shown in solutions of conjugated polymers, however in these studies controlled chemical defects, often in the form of backbone epoxide rings, were introduced to interrupt conjugation [37,38,42].

In this paper, we report experimental findings of photoluminescence quantum yield enhancement from BDMO-PPV in toluene by photochemical reaction. We measure emission, absorbance, reflectance, and transmission of BDMO-PPV in toluene solution under continuous degradation in ambient conditions. The enhancement of both external and internal quantum yield is detailed in the results section.

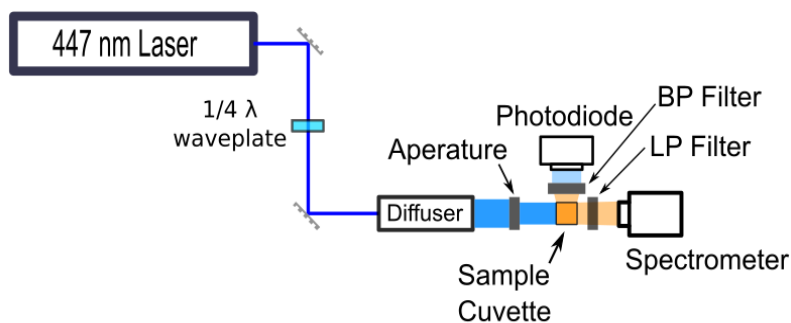

Figure 2 - Experimental setup for photoluminescence and scattered light intensity measurements. BP: bandpass, LP: longpass

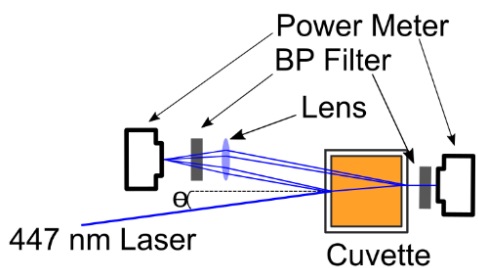

Figure 3 - Experimental setup for direct measurements of reflectance and transmittance. Theta corresponds to a small angle of 0.55°. BP: bandpass

## 2. Experiment

BDMO-PPV (Sigma Aldrich 546518) was purchased from Sigma Aldrich. Solutions of BDMO-PPV were made with room temperature chloroform, chlorobenzene, and toluene in concentrations of 2.5 mg/mL. These high concentration solutions were stored in dark conditions at 4°C to minimize uncontrolled degradation [2]. Solutions for degradation and UV-Vis spectroscopy were made by diluting the original 2.5 mg/mL solutions to 0.0625 mg/mL. These solutions of BDMO-PPV were degraded in a 10 mm fused quartz cuvette using a continuous wave 447 nm laser with a 320 mW/cm$^2$ constant illumination. The laser beam passed through a quarter waveplate and a diffuser to create a spatially homogeneous top-hat beam profile with uniform polarization. Photoluminescence spectra were taken every 30 seconds using an Ocean Optics

FLMT spectrometer with an integration time of 80 milliseconds. The relative intensity of the collected PL spectra was calibrated using a NIST traceable light source. The relative intensity of the scattered light of the excitation laser beam was also collected using a photodiode power meter with a bandpass filter. A schematic experimental setup is shown in Fig. 2.

In a separate setup, small-angle (~ 0.55°) reflectance and transmittance were measured using a low-power laser (2.59 mW at the sample) as shown in Fig. 3 for three states of the sample (pristine, peak QY, and highly degraded). A lens was used to collect all of the reflected beams from the sample cuvette. No lens was needed for the transmission measurement since the photodiode detector area covered all the transmitted beams. The transmittance of the optical components were considered in the calculation. UV-Vis spectroscopy was performed using a Thermo Scientific Evolution 201 UV-Vis spectrometer.

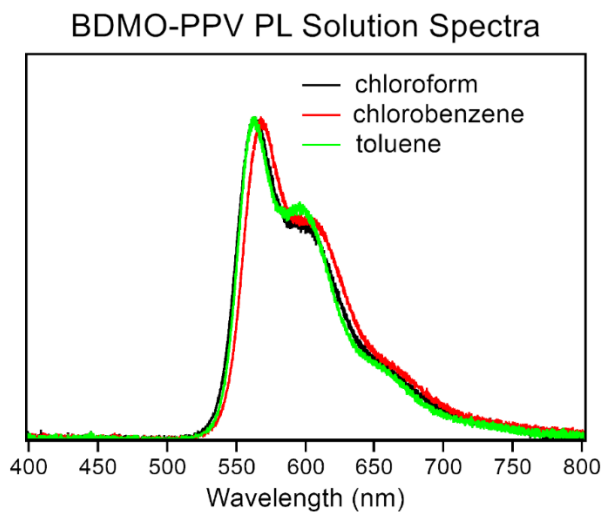

Figure 4 – Normalized photoluminescence spectra from BDMO-PPV in chlorobenzene, chloroform, and toluene.

## 3. Results and Discussion
**External Quantum Yield**

PPV polymers degrade quickly under illumination in oxygen and water environment. We measured the relative number of emitted photons during photodegradation by integrating PL intensity over photon energy E, which is given by

$$N_{photons} = \int \frac{PL(E)}{E} dE \qquad \text{(Eq. 1)}$$

This photon number is proportional to external quantum yield ($QY_{ext}$) defined by the ratio of the number of emitted photons to the number of incident photons.

$$QY_{ext} = \frac{N_{emission}}{N_{incident}} \qquad \text{(Eq. 2)}$$

Photoluminescence emission of PPV backbone polymers can be modified using various solvents such as toluene, chloroform, and chlorobenzene [43,44]. Example PL spectra are shown in Fig. 4 from BDMO-PPV in these three solvents. In Fig. 5, the relative number of PL photons from BDMO-PPV in the three solvents are shown as they degrade under continuous illumination at 447 nm. Similar spectral changes were obtained from previous work [39] and the relative number of the PL photons was calculated using Eq. 1. The number of photons is equivalent to the relative $QY_{ext}$ change in time since the excitation photon flux is constant (Fig. 5). We have found that during the degradation process, $QY_{ext}$ diverges from an exponential decay due to photo-activated intermediate states [39]. Using different solvents affects the amount of quantum yield enhancement as well as the degradation time, showing that this effect is strongly solvent dependent. In particular, BDMO-PPV in toluene showed a significantly long degradation time with the highest $QY_{ext}$ increase during the decay compared to solutions of chloroform and chlorobenzene.

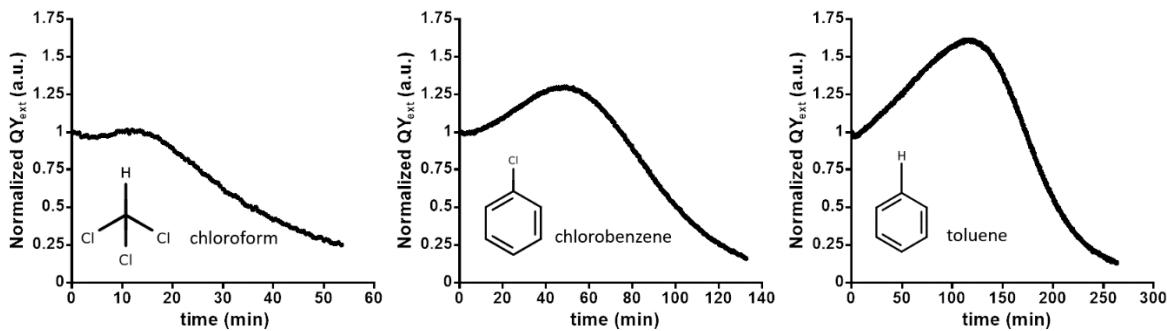

Figure 5 – Relative change of the number of photon emission during BDMO-PPV photodegradation in chloroform, chlorobenzene, and toluene; which is equivalent to external quantum yield change.

**Internal Quantum Yield**

We have also monitored the internal quantum yield ($QY_{int}$) change during the degradation. Internal quantum yield is given by:

$$QY_{int} = \frac{N_{emission}}{N_{absorbed}} = \frac{N_{emission}}{N_{incident} - N_{scattered} - N_{reflected} - N_{transmitted}} \qquad \text{(Eq. 3)}$$

We estimate absorption (A) by measuring scattered light intensity at the excitation beam wavelengths (Fig. 2 and Fig. 6) which is proportional to the scattering rate (S), and reflectance (R) and transmittance (T) for the sample system (Fig. 3). From the relation of A, S, R, and T (Eq. 4), the change of absorption (A) can be determined from the other measurements.

$$A = 1 - (S + R + T) \qquad \text{(Eq. 4)}$$

The intensity of the scattered beam remained near-constant until the $N_{emission}$ ($QY_{ext}$) peak was reached (Fig. 6), and R + T increased from 6.01% to 9.68% (Table 1). Consequently, the absorption of the sample decreased. From S, R, and T measurements, we can conclude that $QY_{int}$ of the sample during photodegradation increased significantly until reaching peak emission.

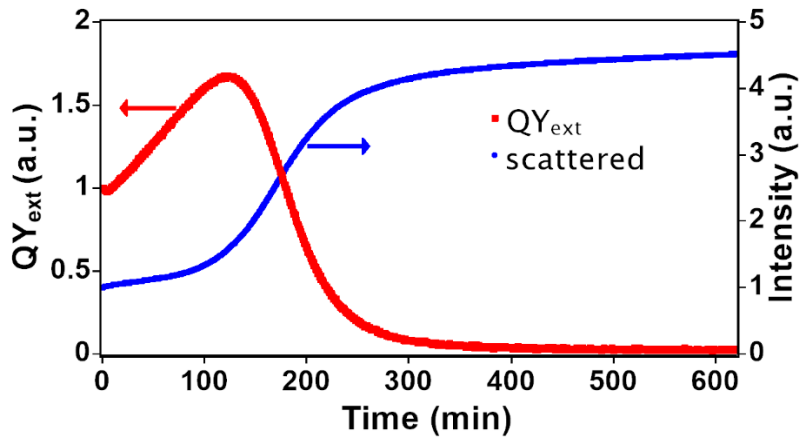

Figure 6 - Integrated number of emission photons, relative QY$_{ext}$, (red) and relative scattered light intensity at the excitation beam wavelength (blue) in BDMO-PPV toluene solution.

|  | Pristine (0 mins) | Peak (120 mins) | Degraded (360 mins) |
| --- | --- | --- | --- |
| Reflectance (R) | 3.11% | 3.12% | 4.66% |
| Transmittance (T) | 2.90% | 6.56% | 74.0% |
| R + T | 6.01% | 9.68% | 78.7% |

Table 1. Reflectance and transmittance values for BDMO-PPV toluene solution.

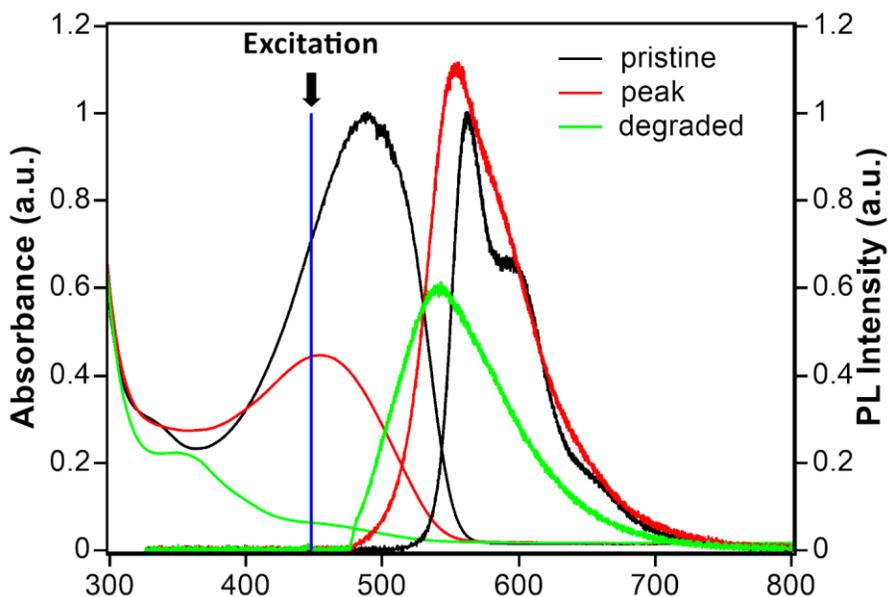

Figure 7 - Absorbance and PL for pristine, peak, and degraded (180 min) BDMO-PPV toluene solution. Excitation wavelength is shown in blue at 447nm.

**Discussion**

Further observations indicate a significant $QY_{int}$ increase at the intermediate state during the photodegradation corresponding to the increase in $QY_{ext}$. From Fig. 7. we compare the absorbance and emission spectra of BDMO-PPV in toluene. The absorbance of the solution decreases over time and the absorbance blue shifts which corresponds to a shorter conjugation length from previous research [22,24,25,27–29,39]. The PL spectrum at the peak of $QY_{ext}$ (red) has 61% more photon emission than the initial PL spectrum as expected from Fig. 5, but has a significantly smaller corresponding absorbance (red). This decrease in absorbed photons is confirmed by reflectance, transmission, and relative scattering measurements using Eq. 4. This demonstrates that during peak $QY_{ext}$ more excitons recombine radiatively rather than non-radiatively, which gives a higher internal quantum yield than pristine BDMO-PPV and suggests that non-radiative recombination sites in BDMO-PPV transform into radiative recombination sites, emitting at shorter wavelengths. A schematic of exciton creation and recombination processes is shown in Fig. 8 for both initial pristine and peak quantum yield states including the creation of blue-shifted highly radiative intermediate states.

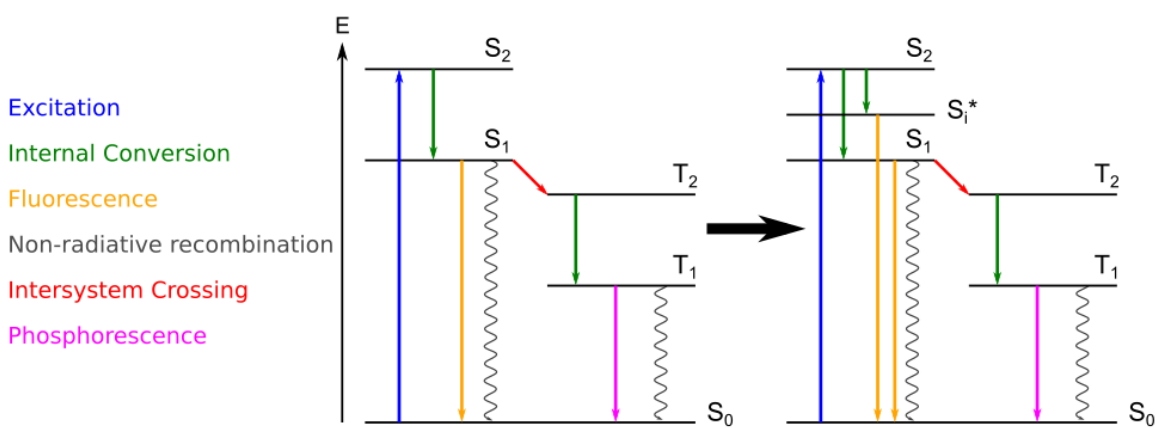

Figure 8 - Jablonski diagrams showing possible electronic states and transition pathways and a new photo-activated state ($S_i^*$) at the peak $QY_{ext}$ shown on the right.

Excitons migrate in luminescent conjugated polymers and recombine radiatively and non-radiatively. During the pristine state, before degradation, there is a mixture of radiative and nonradiative states as shown in Fig 9. As the absorption decreases from the initial pristine state, the amount of radiative states increases, resulting in more photon emission. The effects due to the increase in the number of radiative states are amplified by the simultaneous decrease in total absorption. Previously non-radiative states must transition to radiative sites to allow for the increased emission corresponding to the increase in $QY_{ext}$. After the $QY_{ext}$ peak, $QY_{int}$ decreases monotonically with an increase in scattering and transmission accompanying a decrease in absorption. Our results strongly indicate that during the photodegradation process, non-radiative recombination sites in BDMO-PPV are transformed into highly emissive PL sites (the conversion in Fig. 9), increasing internal quantum yield. This conversion is strongly suggested to be a combination of photochemical and photophysical processes due to the solvent dependence of $QY_{ext}$.

## 4. Conclusion

BDMO-PPV is a photoluminescent semiconducting polymer of the poly-phenylenevinylene family. Unlike MEH-PPV and MDMO-PPV which have asymmetric sidechains, BDMO-PPV has symmetrical sidechains, affecting its properties as well as its degradation mechanism. Using measurements of emission, absorbance, scattering, reflectance, and transmission of dilute BDMO-PPV solutions in toluene, we found photoluminescence quantum yield enhancements during the photodegradation process. We propose a new intermediate emissive state during photodegradation, responsible for an increase in internal and external quantum yield. This intermediate state is long-lived and highly emissive in toluene solutions compared to other solvents, suggesting a mixture of photophysical and photochemical processes during degradation. Further research is required to elucidate the mechanism of the quantum yield enhancements.

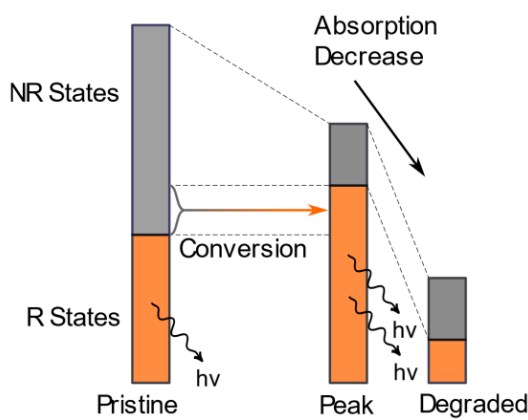

Figure 9 - Diagram showing the conversion of non-radiative (NR) exciton recombination states to radiative (R) states during degradation.